\pdfoutput=1
\RequirePackage[2018-12-01]{latexrelease}
\documentclass{rQUF2e}

\usepackage{epstopdf}% To incorporate .eps illustrations using PDFLaTeX, etc.
\usepackage{subfigure}% Support for small, `sub' figures and tables
\usepackage{graphicx}

\theoremstyle{plain}

\theoremstyle{definition}

\theoremstyle{remark}

\begin{document}

\title{Generation of synthetic financial time series by diffusion models}

\author{
Tomonori Takahashi$^{\ast}$$\dag$\thanks{ $^\ast$ Corresponding author. Email: t\_takahashi@nii.ac.jp} and Takayuki Mizuno${\ddag}$ \\
\affil{
$\dag$ The Graduate University for Advanced Studies, SOKENDAI, 2-1-2 Hitotsubashi, Chiyoda-ku, Tokyo, Japan \\
$\ddag$ National Institute of Informatics, 2-1-2 Hitotsubashi, Chiyoda-ku, Tokyo, Japan
}
%\received{v1.0 released October 2024}
}

\maketitle

\begin{abstract}
Despite its practical significance, generating realistic synthetic financial time series is challenging due to statistical properties known as stylized facts, such as fat tails, volatility clustering, and seasonality patterns. Various generative models, including generative adversarial networks (GANs) and variational autoencoders (VAEs), have been employed to address this challenge, although no model yet satisfies all the stylized facts. We alternatively propose utilizing diffusion models, specifically denoising diffusion probabilistic models (DDPMs), to generate synthetic financial time series. This approach employs wavelet transformation to convert multiple time series (into images), such as stock prices, trading volumes, and spreads. Given these converted images, the model gains the ability to generate images that can be transformed back into realistic time series by inverse wavelet transformation. We demonstrate that our proposed approach satisfies stylized facts.
\end{abstract}

%%%%%%%%%%%%%%%%%%%%%%%%%%%%%%%%%%%%%%%%%%%%%%%%%%%%%%%%%%%%%%%%%%%%%%%%%%%%%%%

\section{Introduction} \label{sec_introduction}

In financial markets, many time series data like stock price fluctuations are generated and recorded every day. Financial time series are of great interest to practitioners and theoreticians for predictions. Because of the enormous amount of data in financial time series and their similarity to physical systems, such as being composed of a large number of agents interacting with financial market participants, financial time series have become a field of study not only in economics but also in statistics and physics \citep{Econophysics, StylizedFacts3}.

It is known empirically that even for such processes as stock prices, which are expected to follow Brownian motion \citep{Bachelier}, the variates follow a fat-tailed distribution rather than a normal distribution. Studies have demonstrated that this distribution obeys a power law and explored the mechanisms by which this power law arises \citep{StylizedFacts1, StylizedFacts2}. Additionally, studies have been conducted on volatility, which represents the magnitude of stock price fluctuations. Attempts have reproduced volatility clustering, which involves persistent periods of high or low volatility, as well as the autocorrelation of the time series of volatilities \citep{StylizedFacts3, StylizedFacts4, StylizedFacts5, ParametricModels1, AgentModel1, AgentModel2}. These properties of financial time series, such as fat-tailed distribution and volatility clustering, are called \textit{stylized facts}. These attempts utilized parametric models, such as the ARCH model and agent-based models, although the former's goal is to represent the mechanism from which stylized facts emerge, not to reproduce plausible financial time series. Through these studies, the existence and universality of stylized facts have been disseminated among researchers.

Another workstream for financial time series is the generation of synthetic time series. Recent remarkable developments in machine learning technology have led to the emergence of models that generate images and natural language texts that resemble those made by human hands. These novel technologies have been applied to the generation of synthetic financial time series data, and some studies suggest the reproduction of realistic financial time series \citep{GAN_TimeSeries_LiteratureReview, GAN_Finance_Overview, QuantGAN, Dogariu}. These studies focused on generating realistic financial time series in terms of reproducing stylized facts. However, despite many efforts, no model has successfully reproduced every stylized fact \citep{Dogariu}.

We address the limitations of the current synthetic financial time series generation methods and propose an approach that employs denoising diffusion probabilistic models (DDPMs) suggested in \citet{DDPM}. Recognized for their excellent image generation, DDPMs surpass generative adversarial networks (GANs) and variational autoencoders (VAEs) in producing high-quality, diverse synthetic data \citep{Trilemma}. Our methodology uniquely incorporates wavelet transformation to convert financial time series into full-color spectrogram images \citep{Wavelet}, enabling DDPMs to learn the intricate characteristics of these series through image analysis. Through post-training, these spectrogram images are transformed back into financial time series data. This method leverages the trichannel nature of color images (RGB) to simultaneously train on and infer three interrelated time series, stock prices, spreads, and trading volumes, all of which are observed within identical periods. We demonstrate that our approach captures these three time series in unison and also accurately reproduces the stylized facts associated with them, marking a significant advancement in the field of financial time series synthesis.

The remainder of the article is organized as follows. In Section 2 we discuss the existing approaches in the literature to the stylized facts of financial time series in econophysics and informatics. Our proposed methodology is presented in Section 3, and Section 4 discusses its experimental setup and results. Finally Section 5 discusses our results and future work.

%%%%%%%%%%%%%%%%%%%%%%%%%%%%%%%%%%%%%%%%%%%%%%%%%%%%%%%%%%%%%%%%%%%%%%%%%%%%%%%

\section{Related works} \label{sec_relatedworks}

Time series data, a critical component for prediction problems, has received considerable attention, especially from numerous studies aimed at enhancing prediction accuracy \citep{Timeseries_forecasting}. Time series data have attracted studies in both econophysics and informatics. The former focuses on the stylized facts of financial time series; the latter informatics concentrates on generating time series data. In this section, we review related works in these fields.

\subsection{Related works in econophysics}

\begin{figure}
\begin{center}
\begin{minipage}{110mm}
  \includegraphics[width=0.9\textwidth]{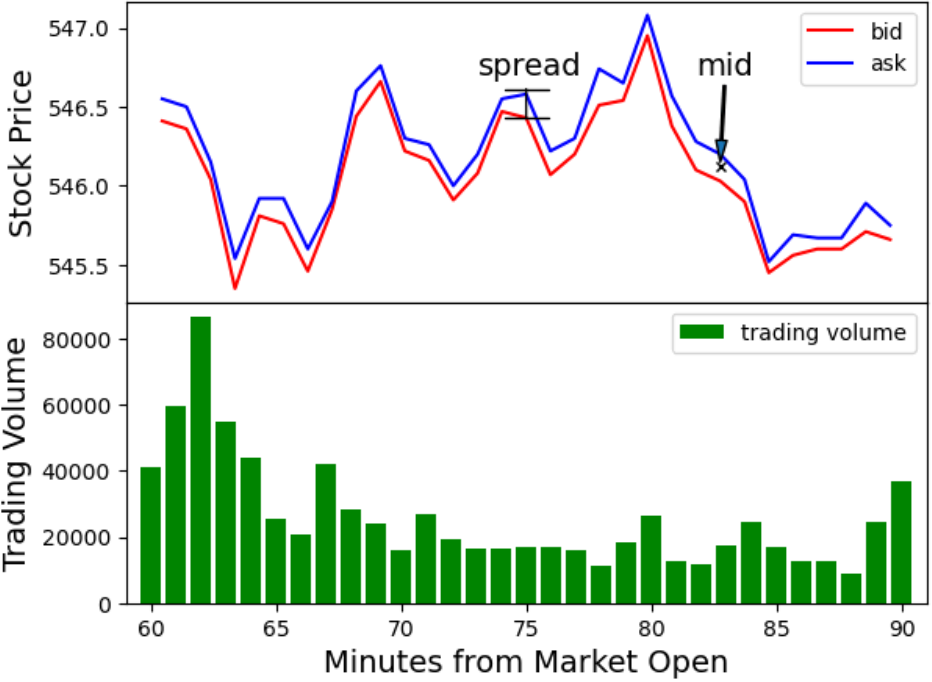}
  \caption{{\bf Examples of typical financial time series.} Data: minute-based stock prices and trading volumes of AAPL.O on NASDAQ on January 23, 2014 from 10:30 to 11:30 AM EST. In a stock exchange, both bid and ask prices are observed. A bid (resp. ask) price is an amount quoted by market participants who want to buy (resp. sell) the stock. Arithmetic average price of bid and ask prices for each time are called the mid-price (the mid), and difference between ask and bid prices is called the bid-ask spread (the spread).}
  \label{Fig1}
\end{minipage}
\end{center}
\end{figure}

Financial time series are generated in actual financial markets every day, every minute, or even more frequently by the activities of the participants in such markets. The most typical example of a financial time series is the price movement of stocks. The spreads and trading volumes of stocks are also observed in financial markets. Here, there are two types of prices: bids and asks. The bid price refers to the buyer's interest price, and  the ask price refers to the seller's interest price. If a market participant wishes to immediately purchase (resp. sell) a particular stock, she can place a request called a market order. As a result, the transaction is executed at the ask (resp. bid) price. The difference of ask and bid prices (generally, bid $<$ ask) is called a bid-ask spread or just the spread. When a transaction is executed by matching a buy market order to the ask price or a sell market order to the bid price, the amount of the transaction's stock is also recorded as time series data, and these amounts are called the trading volumes. The prices and spreads are time series created by observing their snapshots at predetermined intervals: daily, hourly, minute-by-minute, or even more frequently. The trading volumes are also time series based on summing the amounts traded in the intervals. We treat these three typical time series data as financial time series (Fig \ref{Fig1}). Regarding stock prices $S_i$, their log returns $\log \frac{S_i}{S_{i-1}}$ are often used instead of prices to normalize the differences among various stocks. This transformation also moves stock price fluctuations closer to a stationary state, similar to how a random walk, which is a non-stationary time series, often shows stationarity when a difference series is taken.

These financial time series, such as stock prices, resemble Brownian motions. In the early days of theoretical studies of financial time series, modeling was done using Brownian motions \citep{Bachelier}. Over the past few decades, many studies have proved that financial time series exhibit unique characteristics that are not seen in Brownian motions. As explained above, the deviation of financial time series from Brownian motion (called stylized facts) has been studied extensively. One typical stylized fact is the fat-tail phenomenon, where financial time series tend to have heavier tails than a normal distribution, and they follow power laws \citep{StylizedFacts1, StylizedFacts2}. The emergence of these larger than expected movements under a normal distribution is often attributed to the trades of large participants \citep{StylizedFacts1}. Another stylized fact is volatility clustering, where large fluctuations tend to be followed by periods of similarly large fluctuations. This characteristic appears as the slow decay of autocorrelation of volatility time series with respect to lag. Other slow decays of autocorrelations are observed in time series of spreads and trading volumes, which interrelate with volatilities. This characteristic, called the \textit{long memory} of time series, contrasts with the short memory of log returns of stock prices, which have the fast decay of autocorrelation. Such stylized facts are observed not only in stock markets but also in other various financial markets like foreign exchange markets \citep{StylizedFacts6}. Studies have employed parametric models to represent stylized facts. The ARCH model is a key instrument for empirical studies \citep{ParametricModels1}, and some studies have modified random walks \citep{ParametricModels2}. Another approach for studying stylized facts is agent-based models, which simulate market participants as agents \citep{AgentModel1, AgentModel2}.

Additionally, observing the movement of financial time series throughout the day reveals intraday seasonality. Volatility, spread, and trading volume are high immediately after the market opens, decrease during mid-day, and rise again as the market approaches closing. These characteristics have been studied in the context of high-frequency market microstructure \citep{StylizedFacts7}.

Wavelet analysis is another approach for analyzing financial time series \citep{Wavelet}. Wavelet transformation can decompose one-dimensional time series data into two-dimensional pictures in space and time, allowing the nature of the original time series to appear in the pictures.

\subsection{Related works in informatics}

Attempts have explored the nature of time series and the mechanisms that cause them as well as the generation of realistic synthetic time series. For instance, in the medical and social science fields, synthetic time series can support model training through data augmentation while preserving anonymity \citep{GAN_TimeSeries_LiteratureReview}. In the financial sector, there is a practical interest in employing synthetic time series for stress tests under hypothetical market conditions for improving the robustness of risk management and trading models. Given the restricted access to high-frequency financial data and the scarcity of time series that demonstrate such unique phenomena as 'flash crashes,' a concerted effort is focusing on generating diverse, realistic financial time series \citep{GAN_Finance_Overview, QuantGAN, Dogariu}.

Generative adversarial networks (GANs) have been employed to produce not only financial time series but also other general time series. GANs were originally proposed to generate images \citep{Goodfellow}. A GAN has two types of neural network structures: a generator and a discriminator. The generator creates realistic data from random noise data. The discriminator receives real data or data from the generator and decides whether they are real or generated. If the discriminator's decision is correct, the generator incurs a loss; otherwise, the discriminator suffers a loss. By continuing this process as neural network training, the generator eventually learns to generate realistic data. TimeGAN \citep{TimeGAN} is a GAN model specifically designed to generate time series data. It introduces latent space and two functions (embedding and recovery) to map the characteristics of input data into the latent space. As another application of GANs for time series, WaveGAN \citep{WaveGAN} trains GANs by spectrogram images converted from audio data as time series for synthetic audio data generation. For financial time series, QuantGAN \citep{QuantGAN} employs temporal convolutional networks (TCNs) for both its generator and discriminator. As another approach, variational autoencoders (VAEs) \citep{Kingma} are employed for financial time series generation \citep{Dogariu}. A VAE model has latent space and two neural networks, an encoder, and a decoder. The encoder is trained to embed given real data into latent space through the parameters of parametric models, and the decoder is trained to reproduce the given real data. Despite attempts using these generative models like GANs and VAEs to replicate financial time series, no model has yet fully captured all the stylized facts \citep{Dogariu}.

The characteristics denoted as stylized facts complicate the generation of synthetic financial time series. The related works in this area have generally focused on the replication of stylized facts \citep{StylizedFacts3, StylizedFacts4, StylizedFacts5}, including fat tails, volatility clustering, seasonality, and calendar effects. All are absent in Brownian motions. As far as we know, the financial time series focused on in the related works are limited to stock prices, while other time series observed simultaneously with stock prices are out of scope.

In addition to GANs and VAEs, diffusion models are often used to generate images. The critical characteristics of generative models for images are quality, diversity, and the generation speed. The GAN, VAEs, and diffusion models satisfy two of these characteristics, although the third remains challenging. GANs offer quality and speed benefits, VAEs provide strength in diversity and speed, and diffusion models have quality and diversity advantages \citep{Trilemma}. For applications to time series data, diffusion models are used for time series imputation \citep{CSDI}.

%%%%%%%%%%%%%%%%%%%%%%%%%%%%%%%%%%%%%%%%%%%%%%%%%%%%%%%%%%%%%%%%%%%%%%%%%%%%%%%

\section{Methodology} \label{sec_methodology}

As discussed above, building a generative model to generate realistic financial time series remains a challenging task. Our approach for generating synthetic time series data is to utilize a diffusion model: the denoising diffusion probabilistic model (DDPM suggested in \citet{DDPM}). To leverage its advantages (excellent image generation), we convert the time series data into images to train the model. After training, the model generates images, which we then convert back into the original time series data. This approach utilizes the model's  image processing capabilities to handle time series data in a novel and potentially more insightful way. This approach involves the following steps. In the subsequent description, we assume the simultaneous generation of three synchronized observed financial time series: prices, bid/ask spreads, and trading volumes.

\textbf{Preprocessing of financial time series}:
Our methodology for transforming financial time series into synthetic data unfolds through a series of designed steps. 
Initially, because sequence data with size $2^n$ simplify the subsequent discrete wavelet transformation, we expand the time series by mirror expansions at both ends to align the length of the time series to $2^n$.
Then we calculate the log returns of the stock prices by determining the differences of the natural logarithm of consecutive stock prices. This step, which is crucial for addressing the non-stationarity inherent in stock price time series, renders them more amenable to analysis. Concurrently, we apply the $arsinh$ transformation to the trading volumes to approximate a logarithmic scale for large values by and to facilitate effective scale transformation. Unlike the natural logarithm, the $arsinh$ function allows small values near 0 and 0 to enter \citep{Arsinh}.
Following this, we undertake a power transformation on the series and normalize it as $\frac{(X_t - \mu(X_t))^\frac{1}{p}}{\sigma(X_t)}$. The power index can be different for each dimension in multivariate time series.
In the steps so far, time series $X_t$ contains many outliers, which reduce the training and inference efficiency. We substitute such outliers with the given z-values of the normalized data through winsorization. For example, if $X_t > z$ (resp. $X_t < -z$), then such $X_t$ are substituted by $z$ (resp. $-z$).

\begin{figure}
\begin{center}
\begin{minipage}{130mm}
  \includegraphics[width=0.9\textwidth]{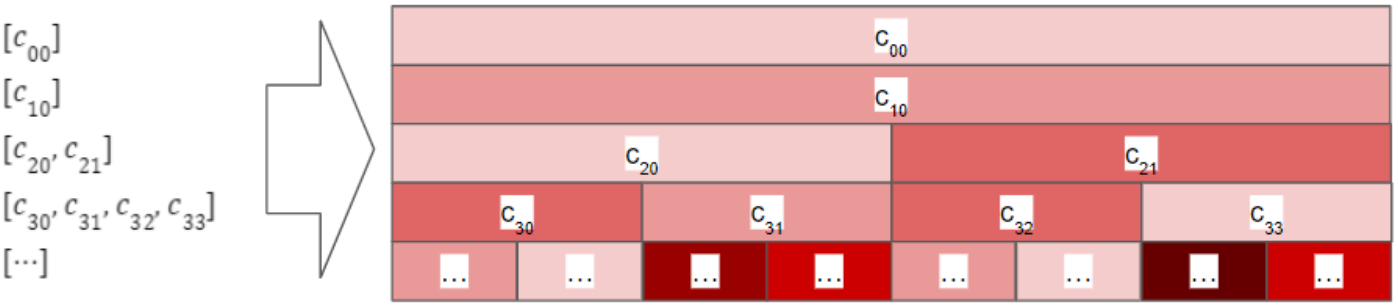}
  \caption{{\bf Pixel imaging.} Wavelet coefficients \\ $c_{00}, c_{10}, c_{20}, c_{21}, c_{30}, c_{31}, c_{32}, c_{33}, \cdots$ are tiled to image pixels.}
  \label{Fig2}
\end{minipage}
\end{center}
\end{figure}

\begin{figure}
\begin{center}
\begin{minipage}{150mm}
  \includegraphics[width=0.9\textwidth]{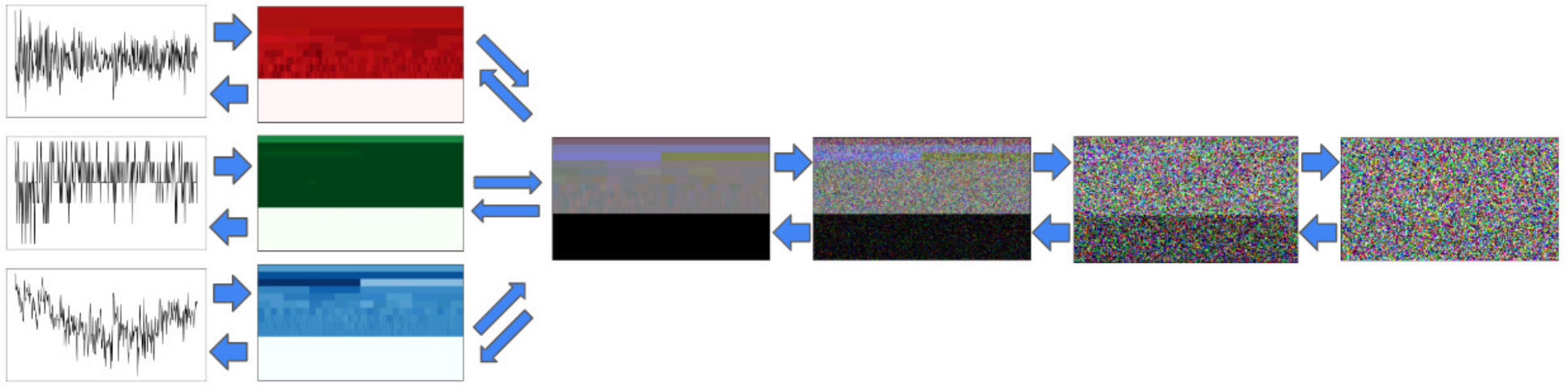}
  \caption{{\bf Overview of the methodology} Pixel filling by wavelet coefficients, overlaying of RGB colors, and noising/denoising of time series images: top left: log returns, middle left: spreads, bottom left: trading volumes.}
  \label{Fig3}
\end{minipage}
\end{center}
\end{figure}

\textbf{Wavelet transformation and filling of pixels of an image}:
After applying discrete wavelet transformation to the preprocessed time series, we obtain sequences of wavelet coefficients. For the transformation, we employed the Haar wavelet as the mother wavelet. Because of the mirror expansion for making the length of the time series $2^n$, we obtain a zero-th order coefficient, a first order coefficient, two second order coefficients, four third order coefficients, and so on up to $(n-1)$-th order coefficients. These coefficients are regarded as luminance and are arranged to fill the pixels, transforming each time series into a monochrome image. In this pixel filling, $k$-th order coefficients are embedded in the $k$-th row of pixels. When the size of the coefficients of each order is $2^l$, the row of pixels is split into $2^l$ areas, each of which is filled with the same coefficient value (Fig \ref{Fig2}). By treating these monochrome images as luminance channels for the three primary colors (red, green, and blue), we synthesize a color image from the trio of monochrome images. The synthesized color image represents a wavelet-transformed time series of price log returns, spreads, and trading volumes. This conversion from three time series data in one interval to a single color image is applied to multiple intervals, resulting in multiple color images as the training dataset. The culmination of this process involves the training of DDPMs using these color images converted from time series data. The synthetic images produced by the trained model are then converted back into time series data, employing the reverse operations of the preceding steps. Fig \ref{Fig3} summarizes these procedures. Through this process, we demonstrate how our approach generates synthetic financial time series data by leveraging the capabilities of DDPMs in learning and reproducing the complex dynamics of financial markets.

\textbf{Model setup}:
DDPMs utilize a UNet architecture to facilitate learning through convolutional processes. This UNet is composed of multi-stage convolutions that include an attention mechanism. For this implementation, we employ the identical channel dimension parameters as found in the Hugging Face tutorial on DDPMs (\texttt{https://huggingface.co/docs/diffusers/tutorials/basic\_training}): 128-128-256-256-512. These parameters define the channel dimensions at various UNet stages, enabling efficient processing and feature extraction across different scales of the input data.

%%%%%%%%%%%%%%%%%%%%%%%%%%%%%%%%%%%%%%%%%%%%%%%%%%%%%%%%%%%%%%%%%%%%%%%%%%%%%%%

\section{Results} \label{sec4_results}

\subsection{Experiments}

We train the diffusion model to output 2-dimensional, 3-channel images that are converted from the time series of price log returns, spreads, and trading volumes by wavelet transformation.

\textbf{Data}:
In our experiment, we selected minute-based stock prices, spreads, and trading volumes of AAPL.O traded on NASDAQ from January 2005 to December 2014. We purchased the data from Refinitiv (LSEG), a financial data provider. The data used in this study consist of minute-level OHLC (Open, High, Low, Close) prices for both bids and asks as well as minute-level trading volumes. During this period, some trading days have minutes without any trades, and we omit such entire trading days in these cases, resulting in 2,481 sample days within this period. One business day opens at 9:30 and closes at 16:00 EST, and we focus on the granularity of these 390 minutes within each trading day. Because the open and close times are fixed during this period, the lengths of one-day time series are identical across the samples. After mirror expansion in preprocessing, the length of the time series becomes 512. Through wavelet transformation and imaging the three time series, the price log returns, the spreads, and the trading volumes, these time series in one day become 16x256 images with three channels.

\textbf{Parameters}:
In preprocessing, we adopt power conversion parameter $p=1.5$ for price log returns and $p=1.0$ for the spreads and trading volumes. The winsorization level is set at 10.0, meaning that outliers beyond $10 \sigma$ are replaced by $10 \sigma$.
The baseline logic is identical to the tutorial implementations on Hugging Face (\texttt{https://huggingface.co/docs/diffusers/tutorials/basic\_training}). We train the model for 100 epochs, and other parameters follow the tutorial for DDPMs on Hugging Face.

\textbf{Computational time}:
Our approach based on DDPM and wavelet transformation takes two hours for training under the above settings and another two hours to generate 2500 images on an NVIDIA GeForce RTX 4090 using the PyTorch framework.

\textbf{Comparison to other methodologies}:
Our methodology compares the outcomes of our DDPMs against established generative models for time series, such as TimeGAN \citep{TimeGAN} and QuantGAN \citep{QuantGAN}. Regarding the DDPM-based approach, we compared it with discrete wavelet transformation and a simple approach, which fills 1x512 matrices with time series data after preprocessing and regards the three 1x512 matrices as one 1x512 color image. The comparison examines the essential stylized facts of financial markets, including their ability to replicate fat-tailed distributions, the slow decay of the autocorrelations of time series of volatilities, spreads and trading volumes, and the characteristic U-shaped pattern of intraday time series. In addition to these examinations, we also investigated the cross correlation coefficients among time series.

\subsection{Evaluation}

\begin{figure}
\begin{center}
\begin{minipage}{90mm}
  \includegraphics[width=0.9\textwidth]{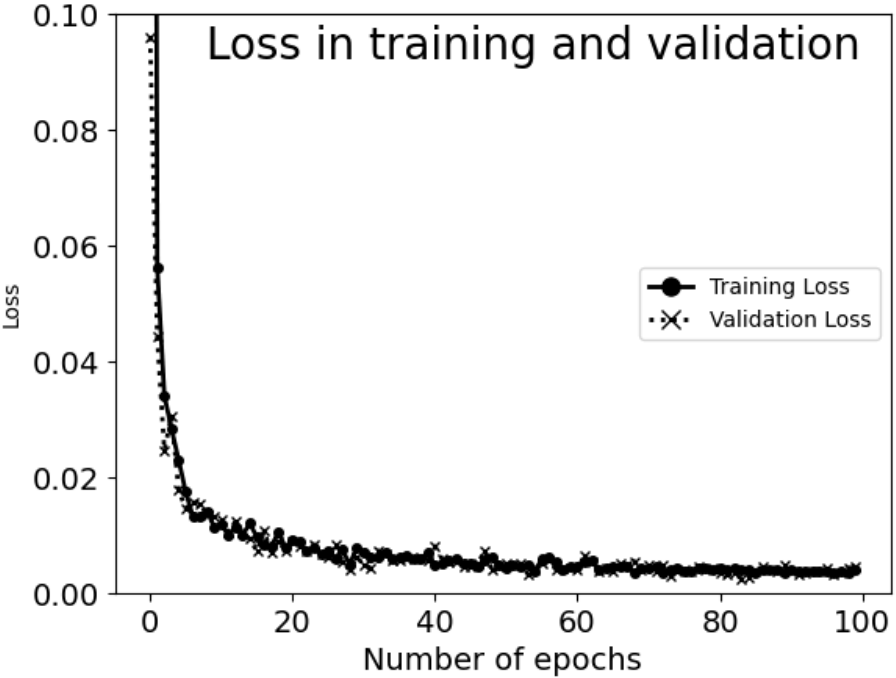}
  \caption{{\bf Losses in training and validation dataset.}}
  \label{Fig4}
\end{minipage}
\end{center}
\end{figure}

\textbf{Losses in training and validation}:
Due to the nature of machine learning models, the values of a loss function decrease with regard to the number of epochs. If the values converge, the model's training works well. In this evaluation, we shuffled the original samples with 2,481 days. 2,000 days of data are regarded as the training dataset and 481 days of data are the validation dataset. Fig \ref{Fig4} explains the convergence of the loss function of the DDPM and the training is processed well.

\begin{figure}
\begin{center}
\begin{minipage}{150mm}
  \includegraphics[width=0.9\textwidth]{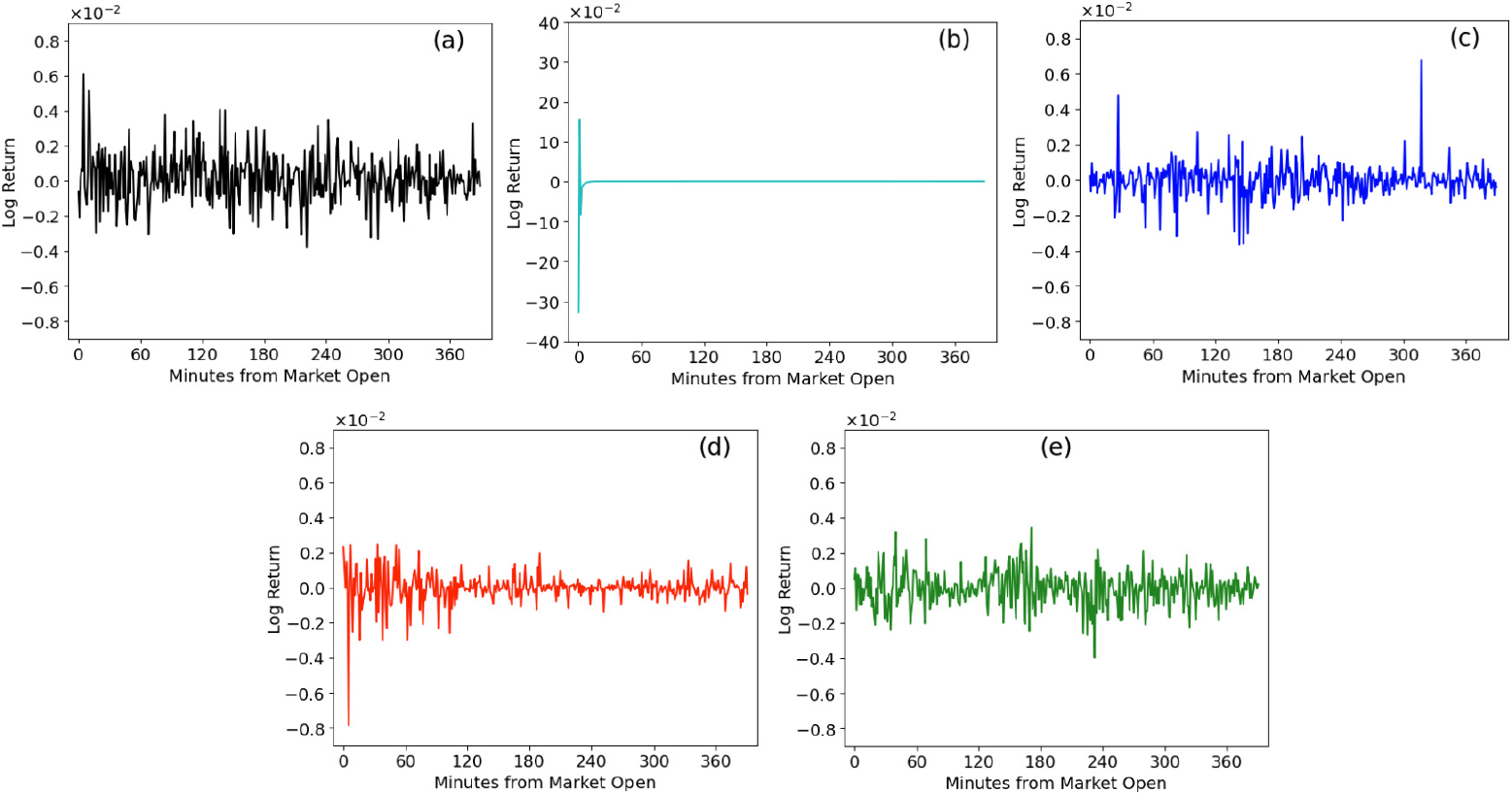}
  \caption{{\bf Shapes of time series of log returns.} \textbf{(a)} real data as of January 7, 2005, \textbf{(b)} synthetic data by TimeGAN, \textbf{(c)} synthetic data by QuantGAN, \textbf{(d)} synthetic data by DDPM with wavelet imaging, and \textbf{(e)} synthetic data by DDPM without wavelet.}
  \label{Fig5}
\end{minipage}
\end{center}
\end{figure}

\textbf{Shape of time series}:
The comparison among TimeGAN, QuantGAN, and DDPMs (both with/without wavelet) reveals that TimeGAN's synthetic time series fail to convincingly replicate the nuanced movements of the log returns of the stock prices, a deficiency that was not rectified by such parameter adjustments as altering the dimensionality of the latent space (Fig \ref{Fig5}). This observation convinced us to focus our comparative analysis on QuantGAN and our DDPM-based approach in terms of log returns. To the best of our knowledge, no prior studies have generated time series data on spreads and trading volumes.

\begin{figure}
\begin{center}
\begin{minipage}{150mm}
  \includegraphics[width=0.9\textwidth]{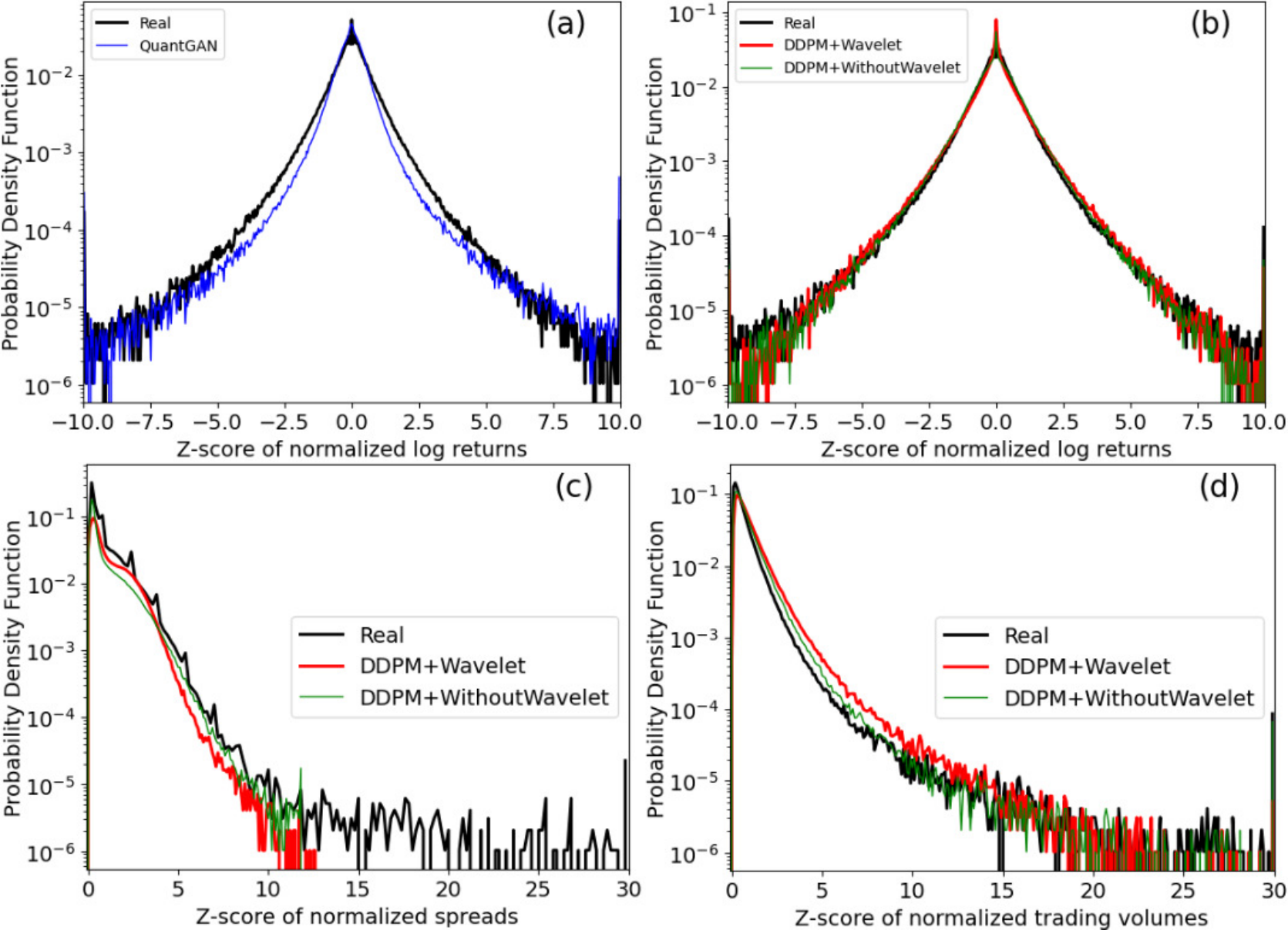}
  \caption{{\bf Probability density functions.} \textbf{(a)} log returns of real data (black) and QuantGAN (blue), \textbf{(b)} log returns of real data (black), DDPM with wavelet imaging (red), and DDPM without wavelet (green), \textbf{(c)} spreads of real data (black), DDPM with wavelet imaging (red), and DDPM without wavelet (green), and \textbf{(d)} trading volumes of real data (black), DDPM with wavelet imaging (red), and DDPM without wavelet (green).}
  \label{Fig6}
\end{minipage}
\end{center}
\end{figure}

\textbf{Fat-tailed distribution}:
The fat-tailed distribution analysis, illustrated in the probability density functions of the log returns (Fig \ref{Fig6}), demonstrates our approach's superiority when mirroring real distributions up to 10$\sigma$. QuantGAN, which uses a 'gaussianizer' in preprocessing to convert the original log return distribution to a Gaussian, shows worse fitting around 1\textasciitilde 2$\sigma$ compared to the real data, although it fits well around 10$\sigma$. Our approach, based on DDPM and wavelet transformation, also shows good fits up to 10$\sigma$ for the spreads and trading volumes.

\begin{figure}
\begin{center}
\begin{minipage}{150mm}
  \includegraphics[width=0.9\textwidth]{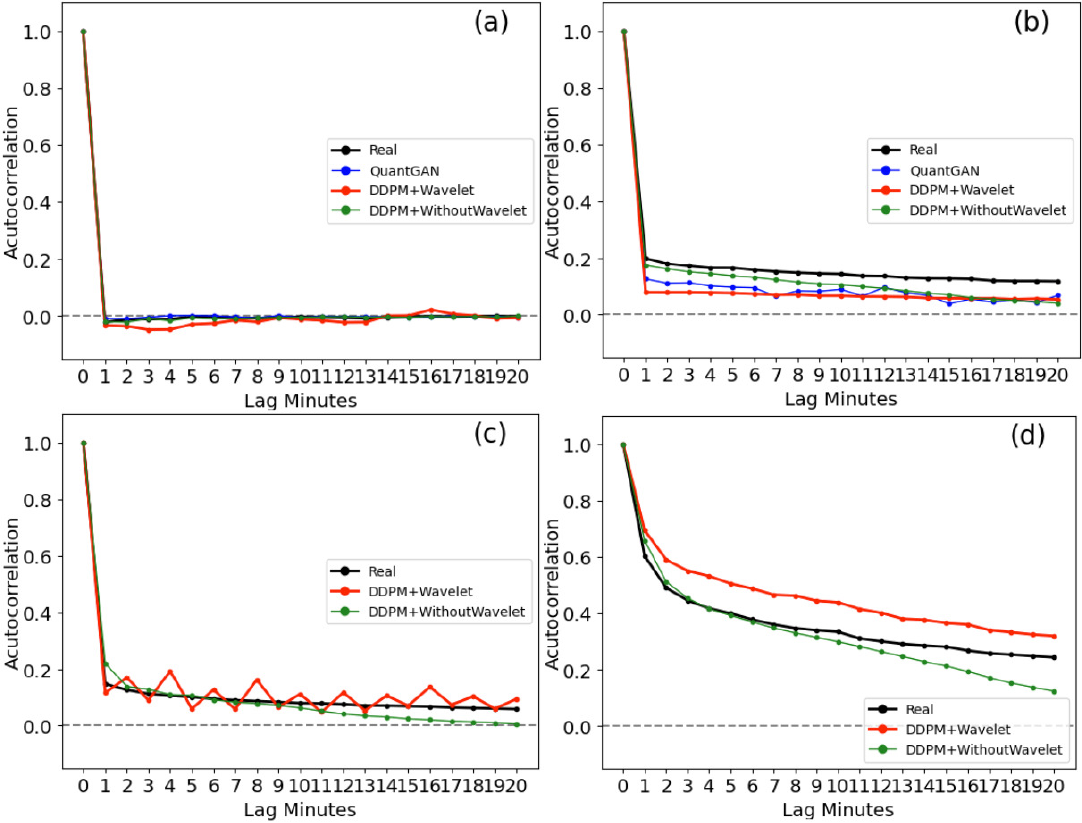}
  \caption{{\bf Autocorrelations.} \textbf{(a)} log returns, \textbf{(b)} volatilities, absolute values of log returns, \textbf{(c)} spreads, and \textbf{(d)} trading volumes. In each chart, black chart represents real data, blue chart represents QuantGAN, red chart represents DDPM with wavelet imaging, and green chart represents DDPM without wavelet.}
  \label{Fig7}
\end{minipage}
\end{center}
\end{figure}

\textbf{Slow decay of autocorrelation}:
Some stylized facts pertain to the features of autocorrelations. In a previous work \citet{StylizedFacts4}, the fast decay of autocorrelations is cited as a feature of log returns, and volatility clustering, which quantifies that high-volatility situations tend to cluster in time, is regarded as positive autocorrelations of volatilities and their slow decay. Here we show that our approach using DDPM and wavelet transformation explains both the fast decay of autocorrelations in the log returns and the positive autocorrelations with the slow decay in volatilities (Fig \ref{Fig7}). In minute-based time series, we also expect similar positive autocorrelation structures and their slow decays in the spreads and the trading volumes. Our approach also explains the existence of such positive autocorrelations and their slow decays.

\begin{figure}
\begin{center}
\begin{minipage}{160mm}
  \includegraphics[width=0.9\textwidth]{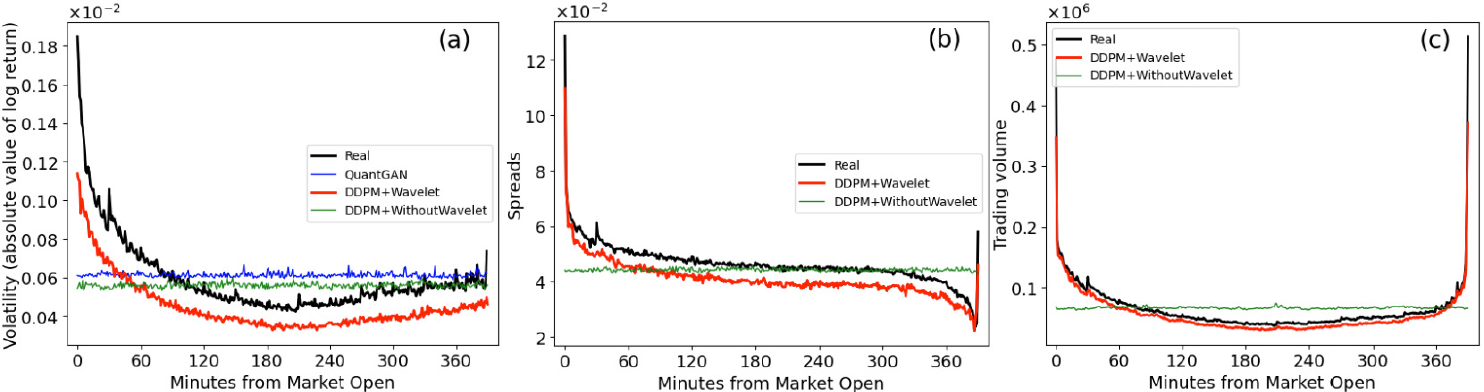}
  \caption{{\bf Intraday seasonality as average of per-minute time series over samples.} \textbf{(a)} volatilities, absolute value of log return, \textbf{(b)} spreads, and \textbf{(c)} trading volume. In each chart, black chart represents real data, blue chart represents QuantGAN, red chart represents DDPM with wavelet imaging, and green chart represents DDPM without wavelet.}
  \label{Fig8}
\end{minipage}
\end{center}
\end{figure}

\textbf{Intraday seasonality}:
Furthermore, our analysis of intraday seasonality, which captures the patterns of the volatilities, the spreads, and the trading volumes, confirms the U-shaped pattern within a day. Fig \ref{Fig8} represents the average taken every minute for the given time series in the real data and for the generated time series. This pattern, starting with high values at the market's opening, dipping mid-day, and rising towards the close, is replicated by our approach, affirming its capacity to accurately mimic real market behaviors. QuantGAN and the simple approach using DDPM without wavelet transformation show flat patterns with regard to time, highlighting their limitations when representing intraday seasonality observed in real markets.

\textbf{Cross correlation coefficients among time series}:
Because our data are synchronized time series observed within a day, we expect them to exhibit cross correlations. Table \ref{Table1} is the cross correlation matrix among time series of real data. A positive correlation coefficient between volatility time series and trading volume time series indicates that high volatility and high trading volume tend to be observed simultaneously. A negative correlation coefficient, between spread time series and trading volume time series, suggests that market participants actively trade higher volumes under tight spreads. Table \ref{Table2} represents the cross correlation matrix among time series generated by DDPM with wavelet imaging, and Table \ref{Table3} shows the cross correlation matrix among time series by simple DDPM without wavelet imaging. Both DDPM-based approaches successfully replicate cross correlation coefficients among time series observed in real data: a positive correlation coefficient between volatilities and trading volumes, a negative correlation coefficient between the spreads and the trading volumes, another negative correlation coefficient between volatilities and spreads, and other quite small correlations.

\begin{table}
\begin{center}
\begin{minipage}{80mm}
\tbl{{\bf Cross correlations among time series of real data}}
{
\begin{tabular}{@{}rrrrr@{}}
\toprule
                & Log Returns & Volatilities & Spreads & Trading Volumes\\
\colrule
Log Returns     & 1 & -0.02 & 0.00  & -0.02 \\
Volatilities    &   & 1     & -0.05 & 0.44  \\
Spreads         &   &       & 1     & -0.13 \\
Trading Volumes &   &       &       & 1     \\
\botrule
\end{tabular}
}
\label{Table1}
\end{minipage}
\end{center}
\end{table}

\begin{table}
\begin{center}
\begin{minipage}{80mm}
\tbl{{\bf Cross correlations among time series of synthetic data (DDPM+Wavelet)}}
{
\begin{tabular}{@{}rrrrr@{}}
\toprule
                & Log Returns & Volatilities & Spreads & Trading Volumes\\
\colrule
Log Returns     & 1 & 0.00  & 0.01  & 0.00  \\
Volatilities    &   & 1     & -0.05 & 0.25  \\
Spreads         &   &       & 1     & -0.12 \\
Trading Volumes &   &       &       & 1     \\
\botrule
\end{tabular}
}
\label{Table2}
\end{minipage}
\end{center}
\end{table}

\begin{table}
\begin{center}
\begin{minipage}{80mm}
\tbl{{\bf Cross correlations among time series of synthetic data (DDPM+WithoutWavelet)}}
{
\begin{tabular}{@{}rrrrr@{}}
\toprule
                & Log Returns & Volatilities & Spreads & Trading Volumes\\
\colrule
Log Returns     & 1 & -0.01 & 0.00  & 0.00  \\
Volatilities    &   & 1     & -0.05 & 0.39  \\
Spreads         &   &       & 1     & -0.14 \\
Trading Volumes &   &       &       & 1     \\
\botrule
\end{tabular}
}
\label{Table3}
\end{minipage}
\end{center}
\end{table}

These findings demonstrate the robustness of our DDPM-based method for generating synthetic financial time series that faithfully reproduce complex market dynamics and stylized facts. These check points are summarized in Table \ref{Table4}. Because TimeGAN failed to replicate the nuanced movements of the log returns of stock prices, we did not check other points. QuantGAN generated only the log returns of the stock prices; the cross correlations among multiple time series were skipped.

\begin{table}
\begin{center}
\begin{minipage}{80mm}
\tbl{{\bf Summary of comparisons among approaches}}
{
\begin{tabular}{@{}rrrrr@{}}
\toprule
                               & TimeGAN & QuantGAN & DDPM (without wavelet) & \textbf{DDPM+Wavelet} \\
\colrule
Shape of time series           & NG      & OK       & OK                     & OK                    \\
Fat tail                       & -       & OK       & OK                     & OK                    \\
Slow decays of autocorrelation & -       & OK       & OK                     & OK                    \\
Intraday seasonality pattern   & -       & NG       & NG                     & OK                    \\
Cross correlation function     & -       & -        & OK                     & OK                    \\
\botrule
\end{tabular}
}
\label{Table4}
\end{minipage}
\end{center}
\end{table}

%%%%%%%%%%%%%%%%%%%%%%%%%%%%%%%%%%%%%%%%%%%%%%%%%%%%%%%%%%%%%%%%%%%%%%%%%%%%%%%

\section{Conclusions}
We suggested an alternative approach to generate synthetic time series by wavelet transformation and denoising diffusion probabilistic model (DDPM). The DDPM and wavelet imaging approaches more effectively replicated characteristics commonly observed in financial time series, such as the fat tails, the slow decay of autocorrelations including volatility clustering, the intraday seasonality patterns, and the cross correlation coefficients among time series, compared to TimeGAN and QuantGAN approaches and the simple application of DDPM without wavelet imaging. The DDPM and wavelet approach especially had a better representation of intraday seasonality in the actual market data than these methodologies in comparison. Imaging through the wavelet transformation can capture intraday seasonality because the relationship among frequencies is more explicit than the original time series. In the preprocessing of imaging, the DDPM with a wavelet imaging approach fills the top rows of an image with a zero-th wavelet coefficient, which represents the overall intraday trend. The next and subsequent pixel rows of the image represent progressively finer market microstructures. As a result, the information of short-term microstructures with multiple time scales by wavelet transformation around market open (resp. close) is embedded in the left (resp. right) side of the image. This contributes to the representation of the U-shape structures of the observed intraday data.

In conclusion, our application of denoising diffusion probabilistic models (DDPMs) in conjunction with wavelet image transformation was an effective method for generating synthetic financial time series that closely adhere to underlying stylized facts. The strategic use of RGB channels in color images to represent and simultaneously generate three interconnected time series replicated the structure of the cross correlation functions among the multiple time series, representing a significant advancement in the field. Looking ahead, the potential to extend this methodology to utilize three or more channels might allow simultaneous generation of multiple correlated stock prices. Building on this foundation, future work is poised to explore the generation of synthetic data that capture even more nuanced relationships and cross correlations within financial markets, extending the boundaries of what is possible in synthetic data generation and financial modeling.

%%%%%%%%%%%%%%%%%%%%%%%%%%%%%%%%%%%%%%%%%%%%%%%%%%%%%%%%%%%%%%%%%%%%%%%%%%%%%%%

%%%%%%%%%%%%%%%%%%%%%%%%%%%%%%%%%%%%%%%%%%%%%%%%%%%%%%%%%%%%%%%%%%%%%%%%%%%%%%%

%%%%%%%%%%%%%%%%%%%%%%%%%%%%%%%%%%%%%%%%%%%%%%%%%%%%%%%%%%%%%%%%%%%%%%%%%%%%%%%%%%%%%%%%%%%%%%%%%%%%%%%%%%%%%%%%%%%%%%%%%%%%%%%%%%%%%%%%%%%%%%%%%%%%%%%%%%%%%%%

\end{document}